\documentclass[10pt,a4paper]{article}
\usepackage{amsmath}
\usepackage{amssymb}
\usepackage{amscd}
\usepackage{times}
\usepackage{graphicx}
\usepackage{euler}
\usepackage{setspace}
\usepackage{endfloat}

\newcommand{\torol}[1]{}

\renewcommand{\emph}{\textbf}

\newcommand{\GA}{\(\mathrm{GABA}_{\mathrm{A}}{\ }\)}
\newcommand{\hide}[1]{}

\title{From Systems Biology to Dynamical Neuropharmacology: Proposal
  for a New Methodology}

\author{P\'eter \'Erdi\\
  Center for Complex Systems Studies, \\ 
  Kalamazoo College, Kalamazoo, Michigan, USA and \\
  Department of Biophysics,\\
  KFKI Research Institute for Particle and Nuclear Physics of the\\
  Hungarian Academy of Sciences, Budapest, Hungary, \\
  \\
  Tam\'as Kiss\\
  Center for Complex Systems Studies,\\
  Kalamazoo College, Kalamazoo, Michigan, USA and \\
  Department of Biophysics, \\
  KFKI Research Institute for Particle and Nuclear Physics of the\\
  Hungarian Academy of Sciences, Budapest, Hungary,\\
  \\
  J\'anos T\'oth\\
  Department of Analysis, Institute of Mathematics, Faculty of Sciences,\\
  Budapest University of Technology and Economics, Budapest, Hungary\\
  \\
  Bal\'azs Ujfalussy\\
  Department of Biophysics,\\
  KFKI Research Institute for Particle and Nuclear Physics of the\\
  Hungarian Academy of Sciences, Budapest, Hungary\\
  \\
  L\'aszl\'o Zal\'anyi\\
  Department of Biophysics,\\
  KFKI Research Institute for Particle and Nuclear Physics of the\\
  Hungarian Academy of Sciences, Budapest, Hungary}

\begin{document}

\maketitle
\doublespacing
\newpage
\section*{Abstract}
\noindent The concepts and methods of Systems Biology are being
extended to neuropharmacology, to test and design drugs against
neurological and psychiatric disorders. Computational modeling by
integrating compartmental neural modeling technique and detailed
kinetic description of pharmacological modulation of transmitter -
receptor interaction is offered as a method to test the
electrophysiological and behavioral effects of putative drugs. Even
more, an inverse method is suggested as a method for controlling a
neural system to realize a prescribed temporal pattern. In particular,
as an application of the proposed new methodology a computational
platform is offered to analyze the generation and pharmacological
modulation of theta rhythm related to anxiety is analyzed here in more
detail.

\newpage

\section{Introduction}
\label{sec:intro}
Systems Biology is an emergent movement to combine system level
description with microscopic details. It might be interpreted as the
renaissance of cybernetics \cite{wiener} and of system theory
\cite{bertal}, materialized in the works of Robert Rosen
\cite{rosen85}. (For an excellent review on applying the system
theoretical tradition to the new Systems Biology see
\cite{wolkenhauer01}.)

To have a system-level understanding of biological systems
\cite{kitano02,csetedoyle02} we should get information from five key
features:

\begin{description}
\item[Architecture.] The structure (i.e.\ units and relations among
  these units) of the system from network of gene interactions via
  cellular networks to the modular architecture of the brain are the
  basis of any system level investigations.
\item[Dynamics.] Spatio-temporal patterns (i.e.\ concentrations of
  biochemical components, cellular activity, global dynamical
  activities such as measured by electroencephalogram, EEG)
  characterize a dynamical system. To describe these patterns
  dynamical systems theory offers a conceptual and mathematical
  framework. Bifurcation analysis and sensitivity analysis reveal the
  qualitative and quantitative changes in the behavior of the system.
\item[Function.] This is the role that units (from proteins via genes,
  cells and cellular networks) play to the functioning of a system
  (e.g.\ our body and mind).
\item[Control.] There are internal control mechanisms which maintain
  the function of the system, while external control (such as
  chemical, electrical or mechanical perturbation) of an impaired
  system may help to recover its function.
\item[Design.] There are strategies to modify the system architecture
  and dynamics to get a desired behavior at functional level. A
  desired function may be related to some ``optimal temporal
  pattern''.
\end{description}

While Systems Biology is now generally understood in a somewhat
restricted way for proteins and genes, its conceptual and mathematical
framework could be extended to neuroscience, as well. Trivially, there
is a direct interaction between molecular and mental levels: chemical
drugs influence mood and state of consciousness. However, ``almost all
computational models of the mind and brain ignore details about
neurotransmitters, hormones, and other molecules.''  \cite{thagard02}.

In this paper we show how to realize the program of Systems Biology in
the context of a new, dynamic neuropharmacology, which was outlined
recently elsewhere \cite{aradierdi}.  Also, we offer a methodology to
integrate conventional neural models with detailed description of
neurochemical synaptic transmission in order to develop a new strategy
for drug discovery.  The procedure is illustrated on the problem of
finding selective anxiolytics.

In particular, our proposed working hypothesis is that\\

\noindent\fbox{%
  \begin{minipage}{\textwidth}
    \begin{itemize}
    \item for given putative anxiolytic drugs we can test their
      effects on the EEG curve by,
    \item starting from pharmacodynamic data (i.e.\ from dose-response
      plots), which of course are different for different recombinant
      receptors,
    \item setting the kinetic scheme and a set of rate constants,
    \item simulating the drug-modulated GABA\,--\,receptor
      interaction,
    \item calculating the generated postsynaptic current,
    \item integrating the results into the network model of the EEG
      generating mechanism,
    \item simulating the emergent global electrical activity,
    \item evaluating the result and to decide whether the drug tested
      has a desirable effect.
    \end{itemize}
  \end{minipage}
}\\

In section \ref{sec:ezvolt} we first review the literature and
existing computer models of anxiety and anxiolytics. Second a review
of GABA\,--\,GABA$_\text{A}$ (ligand\,--\,receptor) interaction is
given with a particular attention on its connection to anxiety
research. In section \ref{sec:ezvan} we elaborate on the proposed
working hypothesis and give an example by integrating a
GABA$_\text{A}$receptor model by Bai et al.\ (1999)\cite{bai} into a
gamma related theta generating interneuron network model by Orb\'an et
al.\ (2001) \cite{orban01}.

\section{Computer models of anxiety}
\label{sec:ezvolt}

\subsection{Biologically motivated modeling of neural circuits}
\label{sec:compartmental}

\subsubsection{Architecture: the septo-hippocampal system}

It was demonstrated (see e.g.\ the seminal book of Gray and McNaughton
\cite{gray00}) that the septo-hippocampal system is strongly involved
in anxiety and related disorders.

\begin{figure}
  \includegraphics[width=\textwidth]{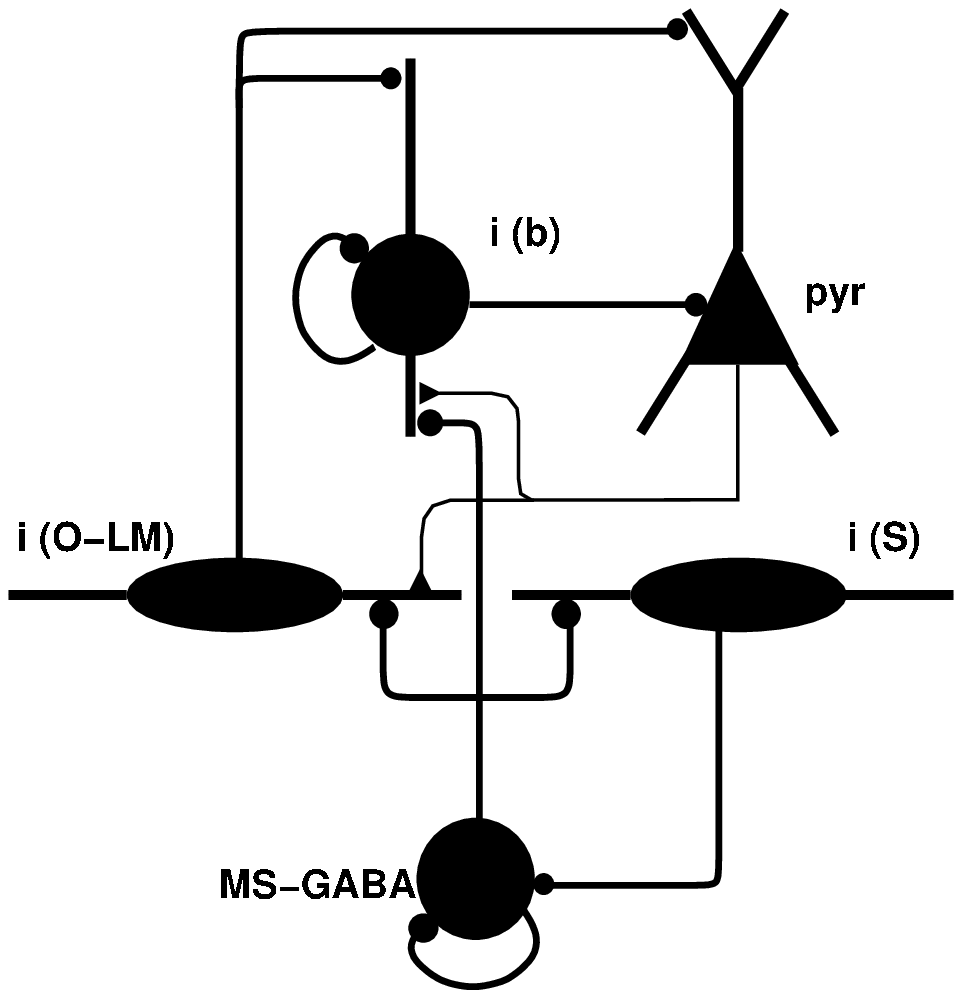}
  \caption[Computer model of the hippocampal CA1 circuitry. Neuron
  populations hypothesized to be responsible for the generation of
  theta oscillation are shown (pyr -- pyramidal cells; i(O-LM) --
  horizontal cells projecting to the distal dendrites of pyramidal
  cells in the lacunosum moleculare layer; i(b) -- basket
  interneurons; i(S) -- septally projecting hippocampal horizontal
  interneurons; MS-GABA -- septal GABAergic cells, triangles denote
  excitatory, dots inhibitory synapses).  Connections originating and
  ending at the same population denote recurrent innervation.]{}
  \label{fig:skeleton}
\end{figure}

In a joint pharmacological and computational work \cite{hajos04,ijis}
effects of the injection of the positive and negative \GA allosteric
modulators diazepam and \mbox{FG-7142}, respectively, were studied.
To investigate the dynamical and functional effects of different
pharmacological agents by computational tools a skeleton model of the
septo-hippocampal system was established. The skeleton network model
(Fig.\ \ref{fig:skeleton}) of the hippocampal CA1 region and the
septal GABAergic cells consisted of five cell populations.  The
hippocampal CA1 pyramidal cell model was a multicompartmental model
modified from \cite{varona00}. In the hippocampal CA1 region basket
neurons and two types of horizontal neurons were also taken into
account.

Connections within and among cell populations were created faithfully
following the hippocampal structure. The main excitatory input to
horizontal neurons is provided by the pyramidal cells via AMPA
(alpha-amino-3-hydroxy-5-methyl-4-isoxazolepropionic acid) mediated
synapses \cite{lacaille87}.  Synapses of the septally projecting
horizontal cells \cite{jinno02} and synapses of the other horizontal
cell population, the O-LM cell population innervating distal apical
dendrites of pyramidal cells \cite{lacaille90} are of the \GA type.
O-LM neurons (on of the two horizontal interneurons) also innervate
parvalbumin containing basket neurons \cite{katona99}.  Basket neurons
innervate pyramidal cells at their somatic region and other basket
neurons \cite{freund96} as well.  Septal GABAergic cells innervate
other septal GABAergic cells and hippocampal interneurons
\cite{freund88,varga02} (Fig.\ \ref{fig:skeleton}). For a full
description of this model see the online supplementary materials to
the paper \cite{hajos04} at:
\mbox{\texttt{http://geza.kzoo.edu/theta/theta.html}}.

The above described model captures several elements of the complex
structure of the hippocampal CA1 and can be used to account for very
precise interactions within this region. However, when the focus of
interest is rather on general phenomena taking place during rhythm
generation modelers might settle for a simpler architecture. In
\cite{orban01} authors describe gamma related theta oscillation
generation in the CA3 region of the hippocampus. The architecture of
the model is exceedingly simplified: only an interneuron network is
simulated in detail. This simplification, however, allowed the authors
to introduce an extrahippocampal input and study its effect on rhythm
generation. As a result, the model is able to account for basic
phenomena necessary for the generation of gamma related theta
oscillation. As an extension of this model, authors show \cite{kiss01}
that activity of the interneuron network indeed accounts for rhythm
generation in pyramidal neurons.

\subsubsection{Dynamics: Generation of theta rhythms}

Theta frequency oscillation of the septo-hippocampal system has been
considered as a prominent activity associated with cognitive function
and affective processes. To investigate the generation and control of
theta oscillation in the hippocampal CA1 region the previously
described detailed, realistic model \cite{hajos04} was used. Firing of
neurons of the four simulated populations were modulated in time.
There are time intervals in which firing was significantly reduced
were alternated by intervals where enhanced firing was observed. This
synchronized state of neural firing was further confirmed by the field
potential, which exhibited a prominent $\approx$5~Hz oscillation (see
Fig.\ 6 in \cite{hajos04}).

Simulation results showed that key components in the regulation of the
population theta frequency are the membrane potential oscillation
frequency of pyramidal cells, strength of the pyramidal cell\,--\,O-LM
cell innervation and the strength of recurrent basket cell
connections.  Amplitude and frequency of pyramidal cell membrane
potential oscillation is determined by their average depolarization,
passive membrane parameters and parameters of the active currents.
Average depolarization in our model results from septal cholinerg
innervation.  An important factor is the presence and maximal
conductance of the hyperpolarization activated current. If
$I_\text{h}$ is present it shortens response times of pyramidal cells
to hyperpolarizing current pulses and more importantly decreases its
variance: $I_\text{h}$ acts as a frequency stabilizer. Synaptic
strengths in our description are set by convergence numbers and
maximal synaptic conductances.

An explanation of intrahippocampal theta oscillation generation --
based on this model -- includes (i), signal propagation in the
pyramidal cell $\to$ O-LM cell $\to$ basket cell $\to$ pyramidal cell
feed-back loop, (ii), synchronization of neural activity via the
recurrent, inhibitory \GA connections within the basket cell network
and (iii), synchronization of pyramidal cell firing due to rebound
action potential generation. It is true that the propagation of a
single signal throughout this trisynaptic loop would not require the
amount of time characteristic to the theta oscillation
($\approx$0.2--0.25~sec), thus in the present case the population
oscillation is created not by the propagation of single signals but
rather the propagation of a ``synchronized state'' in the network.

The observed periodic population activity is brought about by
alternating synchronization and desynchronization of cell activities
due to the interplay of the above mentioned synchronizing forces and
some desynchronizing forces (such as heterogeneity of cell parameters
and diversity of synaptic connections), as observed in
\cite{orban01,kiss01}.

\subsubsection{Function: Mood Regulation}

The hippocampal formation is known to be involved in cognitive
processes (navigation and memory formation -- for reviews see e.g.\
\cite{lengyel05,buzsaki05}) and mood regulation
\cite{gray00,freund03}. When an impairment occurs in the architecture
or dynamics of the septo-hippocampal system, this change is reflected
in its function e.g.\ as an impairment of its memory storing or mood
regulatory capability. One such mood disorder is anxiety:

``Anxiety is a complex combination of the feeling of fear,
apprehension and worry often accompanied by physical sensations such
as palpitations, chest pain and/or shortness of breath. It may exist
as a primary brain disorder or may be associated with other medical
problems including other psychiatric disorders\dots

A chronically recurring case of anxiety that has a serious affect on
your life may be clinically diagnosed as an anxiety disorder. The most
common are Generalized anxiety disorder, Panic disorder, Social
anxiety disorder, phobias, Obsessive-compulsive disorder, and
post-traumatic stress disorder'' \cite{wikipedia}

It is well documented that anxiolytics and hypnotics reduce the
amplitude of septo-hippocampal oscillatory theta activity, which
contributes to their therapeutic effect but causes unwanted side
effects, e.g.\ cognitive impairment as well \cite{maubach04,lees04}.

Historically used mood regulators act on the barbiturate or
benzodiazepine (BZ) sites of GABA receptors. These drugs have both
anxiolytic and hypnotic activity because they modulate all types of
\GA receptors.

Conventional anti-anxiety drugs have sedative-hypnotic side effects,
and have negative effects on cognitive functions, such as memory and
generally they reduce the amplitude of theta rhythm.  While it seems
possible to dissociate sedative-hypnotic effects and anxiety, it is
not clear how to decrease anxiety without impairing memory function.\\

\noindent\textbf{Models of anxioselective actions: search for data}

\noindent Recently it became clear that $\alpha$ subunits of
GABA$_\text{A}$ receptors exhibit a
remarkable functional specificity.  Genetic manipulations helped to
show that $\alpha_1$ subunits are responsible for mediating sedative
effects, while $\alpha_2$ subunits mediates anxiolytic effects
\cite{rud-moehler04}.  Preliminary experimental data and modelling
studies for the effects of the preferential \GA $\alpha_1$ and
$\alpha_2$ positive allosteric modulator, zolpidem and \mbox{L838,417},
respectively, for the septo-hippocampal theta activity have been
reported \cite{balazs-mitt05}.

In that study we examined the effects of the $\alpha_1$ and $\alpha_2$
subtype-selective BZ site ligand zolpidem and \mbox{L838,417} on the
septo-hippocampal system.  In electrophysiological experiments
extracellular single unit recordings were performed from medial
septum/diagonal band of Broca with simultaneous hippocampal (CA1)
electroencephalogram (EEG) recordings from anesthetized rats.  Both of
the drugs eliminated the hippocampal theta oscillation, and turned the
firing pattern of medial septal cells from periodic to aperiodic, but
only the zolpidem reduced the firing rate of the these neurons. In
parallel to these experimental observations, a computational model has
been constructed to clearly understand the effect of these drugs on
the medial septal pacemaker cells.  We showed that the aperiodic
firing of hippocampo-septal neurons can reduce the periodicity of the
medial-septal cells, as we have seen in the case of the
\mbox{L838,417}.  The reduction of firing rates in the case of
zolpidem is attributed to the increase of the synaptic conductances
and the constant inhibition of these cells.  We modelled these drug
effects by modifying (i) the synaptic maximal conductances of the GABA
synapses.  (ii) the constant excitatory drive of the median septal
cells and (iii) the hippocampal input.  The incorporation of a more
detailed synaptic model, similarly to the one proposed in Section
\ref{sec:model}, taking into account differences between $\alpha_1$
and $\alpha_2$ subunits is in progress.

Zolpidem increases by concentration-dependent manner the duration and
amplitude of the postsynaptic current, most likely by enhancing the
affinity of the receptors for GABA \cite{perr-rop99}, but these
effects were diminished or absent in neurons from $\alpha_1$ knock-out
mice \cite{goldsteinetal02}

There seem to be compounds, which might have comparable binding
affinity but different efficacies at the various subtypes, thereby
preferentially exerting their effects at subtypes thought to be
associated with anxiety. \mbox{L838,417} seems to be an example for
subtype selective compounds \cite{atack}, but only few kinetic or even
pharmacodynamic data could be found in the public domain. In Section
\ref{sec:ezvan} drugs whose effect on the GABA kinetics is known will
be used to demonstrate the proposed methodology, using a detailed
kinetic model of GABA receptors within the network model of CA3
interneurons.

\subsection{Phenomenological and kinetic descriptions of synaptic
  interaction}

\subsubsection{Control: GABA receptor kinetics}

\noindent\textbf{Pharmacological modulation: why and how?}

\noindent One perspective of the pharmacotherapy of anxiety might be
based on reducing septo-hippocampal theta activity and it is related
to modulate the interaction between the GABA transmitter and \GA
receptors.  Since \GA receptors occur in a variety of specific forms,
they might be subject of pharmacological control to obtain selective
functional effects. While by-and-large it is understandable that
putative drugs induce opening and closing of synaptic channels (here
the chloride channels), the underlying finely tuned kinetic mechanism
of the interaction of transmitters, modulatory agents and receptors
are not known.  Our longer term goal is to find out how
pharmacological effects depend on the concentration of transmitter and
of modulators. In technical terms, while there are detailed models of
the presynaptic transmitter release here we restrict ourselves to
build a detailed kinetic model for the generation of the postsynaptic
response.\\

\noindent \textbf{Pharmacological elements}

\noindent \textbf{Receptor Structure:} \GA receptors are pentameric
structures consisting of multiple subunits. At this moment
\cite{farr-nus05} nineteen subunits have been cloned from mammalian
brain. According to their sequence similarities, subunits have been
grouped into seven families: $\alpha, \beta, \gamma, \delta, \epsilon,
\pi$ and $\theta$.  Only a few dozen among the many combinatorial
possibilities exist. The most frequent subtypes are two $\alpha$, two
$\beta$ and one $\gamma$ subunits. The structural variations imply
functional consequences \cite{farr-nus05}, among others for the
kinetic properties.\\

\noindent \textbf{Pharmacodynamics:} Effect of drugs can be
characterized by dose-response curves. Some drugs have the ability to
open a certain ion channel per se. In this case drugs are
characterized by the maximal evoked response or the potency, the drug
concentration, which elicit 50~\% of the maximal response.

There are other cases, however, when the drug can not open the ion
channel directly but requires an additional ligand to evoke an action.
In the followings we study drugs that modulate interaction between the
GABA transmitter and \GA receptors.

Both barbiturates and BZs shift the GABA concentration-response curve
to the left, but barbiturates also increase the maximum response. They
act on different states, consequently they have different kinetic
effects: average open time of the channel, but not the channel opening
frequency is increased significantly by barbiturates. As opposed to
BZs, barbiturate receptors do not contain $\gamma$ subunits (see
later).  One more difference is that at high concentration GABA
receptor channels can directly be opened by barbiturates. For a
summary see \cite{martin}.  Anxiolytic activity was not a particular
disadvantage when these drugs were used as hypnotics, hypnosis was a
definite disadvantage when they were used as anxiolytics.  Recent
discoveries made possible the separation between hypnotic and
anxyolitic activity and selective hypnotic agents (e.g.\ zolpidem) are
already on the market. Selective anxiolytics are on the preclinical
and/or in clinical trial stage.\\

\noindent \textbf{The conventional tool of computational neuroscience: a
  phenomenological model}

\noindent One way to describe synaptic transmission is to use a gating
variable similar to the well known Hodgkin\,--\,Huxley formalism:

\begin{subequations}
  \begin{align}
    I_\mathrm{syn} &= \bar{g}_\mathrm{syn} s (V -
    E_\mathrm{syn})\label{eq:isyn}\\
    \frac{\mathrm{d}s}{\mathrm{d}t} &= \alpha
    F\left(V_\mathrm{pre}\right) \left(1 - s \right) - \beta s\\
    F\left(V_\mathrm{pre}\right) &= \frac{1}{1+
      \exp\left(\displaystyle
        \frac{V_\mathrm{pre}-\Theta_\mathrm{syn}}{K}\right)}
  \end{align}
\end{subequations}

\noindent with $I_\mathrm{syn}$ being the synaptic current,
$\bar{g}_\mathrm{syn}$ the maximal synaptic conductance, $s$ the
gating variable of the synaptic channel, $E_\mathrm{syn}$ the synaptic
reversal potential, $F(\cdot)$ is an activation function, $\alpha$ and
$\beta$ rate functions describing opening and closing of the gate of
the synaptic channel, $\Theta_\mathrm{syn}$ is a threshold.

\begin{figure}
  \includegraphics[width=\textwidth]{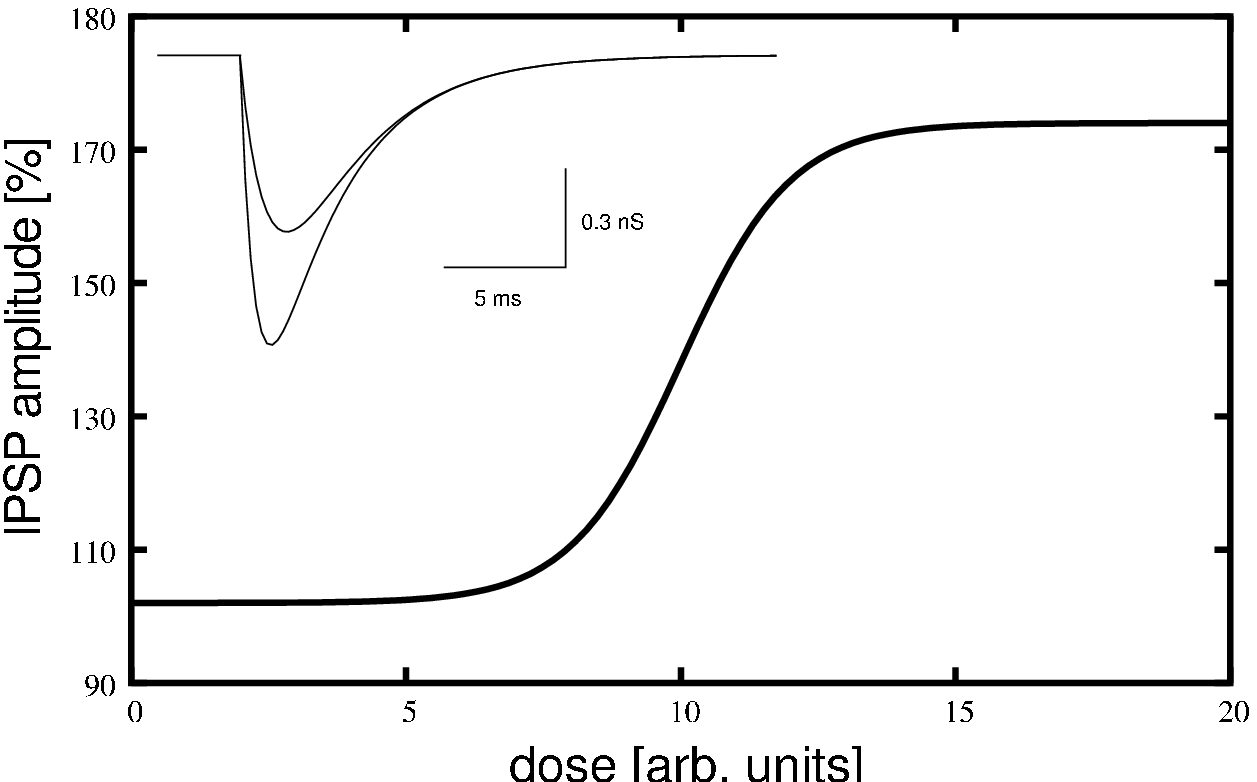}
  \caption[Modelling the effects of allosteric GABA$_\text{A}$
  receptor modulators. In a simple description of synaptic transfer
  the strength of synapses was modulated via the $\bar{g}_\text{syn}$
  parameter in eq.\ (\ref{eq:isyn}) in a dose dependent manner.
  \emph{Inset}: modelled inhibitory postsynaptic potentials before
  (smaller amplitude) and after (larger amplitude) administration of
  positive GABA$_\text{A}$ allosteric modulator.]{}
  \label{fig:dose}
\end{figure}

Figure \ref{fig:dose}.\ illustrates the general form of effects of \GA
receptor modulators described by the phenomenological model, where the
effect of drugs was taken into account solely by modifying
$\bar{g}_\mathrm{syn}$, the maximal synaptic conductance.

\noindent \textbf{From pharmacodynamics to detailed kinetic schemes}

\noindent Kinetic models of transmitter-receptor interaction (without
and in the presence of drugs) help to calculate the induced synaptic
current, and gives a possibility to get quantitative insight how to
modulate it.  A specific step in the transmitter-receptor interaction,
namely desensitization, plays an important role in shaping the
GABAergic currents. Since our general goal is to incorporate
appropriate kinetic models to the network model to compute the drug
effects on global brain rhythm, we should have appropriate kinetic
model.

As opposed to the conventional phenomenological description, a more
effective, but certainly most expensive, modelling tool to evaluate
the pharmacological effects of the different modulators, or even to
give help for offering new putative molecules for drug discovery, is
the inclusion of more detailed kinetic studies of GABA receptor
modulation.

Jones and Westbrook \cite{joneswester95} established a model for
describing the rapid desensitization of the \GA receptors. More
specific kinetic models should be studied to describe the effects of
the different (full and partial) agonists and antagonists taking into
account that connections between singly and doubly bound open and
desensitized states may influence the receptor kinetics. Baker et al.\
\cite{bakeretal02} explained the functional difference between the
effects of propofol (which has hypnotic effect) and of midazolam (a
sedative - amnestic drug) based on a detailed kinetic model of a
single cell with autapses and two interconnected cells.

The main difference is that propofol modifies the desensitization
processes, more dramatically the slow desensitization steps and the
modified kinetic parameters. These differences imply distinct behavior
of the network (synchronization, frequency of oscillation) and
therefore also in function.

\section{Integrating receptor kinetics and neural network models}
\label{sec:ezvan}

\subsection{Setting up the model}
\label{sec:model}

In the following section we connect the two concepts reviewed in the
first part of this paper. Below is the description of an illustrative
model integrating the detailed kinetic description of synaptic
receptors \cite{bai} into the biophysical model of a gamma related
theta rhythm generating interneuron network \cite{orban01,kiss01}. We
use this model to study the effect of drugs that have a well
identified effect on the chemical reactions taking place in the
GABA$_\text{A}$ receptor.

Here, the model is introduced only briefly, detailed equations and
parameters can be found in the Appendix. The modeled interneurons
contained the Hodgkin\,--\,Huxley channels (sodium, potassium and
leak), which were identical for all cells (see \ref{eq:dv}\,--\,
\ref{eq:betan}). Also, the synaptic current ($I_\text{syn}(t)$) and an
input current ($I_i(t)$) were considered (see below).

A connection between two interneurons was established through a
synaptic mechanism consisting of a phenomenological presynaptic and a
chemically realistic described postsynaptic part. The presynaptic model
describes the transmitter release due to an action potential generated
in the presynaptic neuron by a sigmoid transfer function (see
\ref{eq:presyn}). The value of the released transmitter concentration
($[L]$) is then used in the postsynaptic GABA$_\text{A}$ receptor
model to calculate the concentration of receptor proteins being in the
conducting (two ligand binded open) state ($[L2O](t)$) (see
\ref{basicmodel}\,--\,\ref{del2o}). The synaptic current was thus
given by

\begin{equation}
  I_\text{syn}(t) =
  \sum_{i=1}^N\bar{g}_{\text{syn},i}\cdot[L2O]_i(t)\cdot(V(t)-V_\text{syn})
\end{equation}

\noindent where $N$ is the number of presynaptic neurons, $i$ indexes
these presynaptic neurons, $\bar{g}_\text{syn}$ is the maximal
synaptic conductance representing the synaptic channel density, $V(t)$ is
the postsynaptic membrane potential and $V_\text{syn}=-75\,mV$ is the
reversal potential of the chloride current. For details and parameters
of the postsynaptic model see the Appendix.

Using this synaptic mechanism interneurons were connected into a
random network. Probability of forming a connection between any two
neurons was from 60\,\% to 100\,\% in the simulations; autapses were
not allowed. Maximal synaptic strength was normalized to enable
scalability of the model: for each postsynaptic neuron
\mbox{$\sum_{i=1}^N\bar{g}_{\text{syn},i} = \bar{g}_\text{syn}$}, where
$\bar{g}_\text{syn}$ was varied from 0.1\,mS/cm$^2$ to 0.2\,mS/cm$^2$.

As it was shown in previous computer models \cite{orban01,kiss01} when
the interneurons are driven by periodically modulated input and the
phase of inputs to different cells is heterogeneous, the random
network of interneurons generate gamma related theta frequency
resonance. Indeed, as shown before by Wang and Buzs\'aki (1996)
\nocite{wang96} a network of mutually interconnected fast firing
inhibitory neurons can generate synchronized gamma frequency
oscillation. This synchronization is reflected e.g.\ in the population
activity or in the raster diagrams (for these measures see the
Appendix). However, when cells are driven by a periodic inputs with
disperse phase instead of a constant current, a slow (theta frequency
range) modulation of the population activity emerges (for details on
necessary and sufficient conditions for theta generation the reader is
kindly directed to Orb\'an et al.\ (2001)\nocite{orban01}). Briefly,
the input for each cell took the following form:

\begin{equation}
  I_i(t)= DC\cdot\sin(\omega t+\phi_i)+ DC
\end{equation} 

\noindent with $DC$ being the amplitude and the offset of this current
(0.005~--~0.01~A/m$^2$), $\omega$ the frequency (25~--~50~Hz) and
$\phi_i$ is the phase drawn from a Gaussian distribution with 0 mean
and 30~degree variance.

\subsection{Direct problem: in silico drug screening with the
  integrated model}
\label{sec:direct}

Kinetic modeling of synaptic transmission has a flexibility in the
level of detailed description from chemical kinetic to simplified
representation \cite{dest98}. The development of new pharmacological,
electrophysiological and computational techniques make possible to
investigate the modulatory effects of putative drugs for synaptic
currents, and consequently for local field potentials and even
behavioral states. Putative drugs with given kinetic properties can be
tested in silico before (instead of?) real chemical and biological
studies.

As noted before population oscillations reflected by the EEG can be
considered as biomarkers for some brain disorders. We used our
integrated model to study effects of two known drugs, the hypnotic
propofol and the sedative midazolam on the population theta rhythm. In
a previous work \cite{kiss01} it was shown that the population
activity of the interneuron network can be used to account for the
activity of pyramidal cells, which can be in turn translated into EEG.
Thus, to quantify our results the raster plots, a population activity
measure and its Fast Fourier Transform were used (see Appendix).

\textbf{Control condition:} As shown previously \cite{orban01,kiss01}
in control situation this model exhibits spike synchronization in the
gamma frequency band, which is modulated in the theta band
(Fig.~\ref{fig:control}) (for a detailed explanation of measures used
below see the Appendix). On Fig.~\ref{fig:control}\textit{A} (the
raster plot) each row symbolizes the spike train of a given cell with
a black dot representing the firing of an action potential. Note that
firing of the cells line up in well-synchronized columns e.g.\ from
approximately 2.2~sec to 2.35~sec.  As a result of heterogeneity in
the phase of the driving current and in synaptic contacts this high
synchrony loosens, column on the raster plot are less well organized
in the 2.35~sec\,--\,2.45~sec interval. This
synchronization\,--\,desynchronization of cells is a result of the
modulation of the instantaneous frequency of individual cells (Fig.\
\ref{fig:control}\textit{B} -- for more details the reader is kindly
referred to Orb\'an et al (2001)\nocite{orban01}). A qualitative
measure of the synchrony is given by the population activity
(Fig.~\ref{fig:control}\textit{C}) and the network synchrony
(Fig.~\ref{fig:control}\textit{D}).

\begin{figure}
  \includegraphics[height=\textheight]{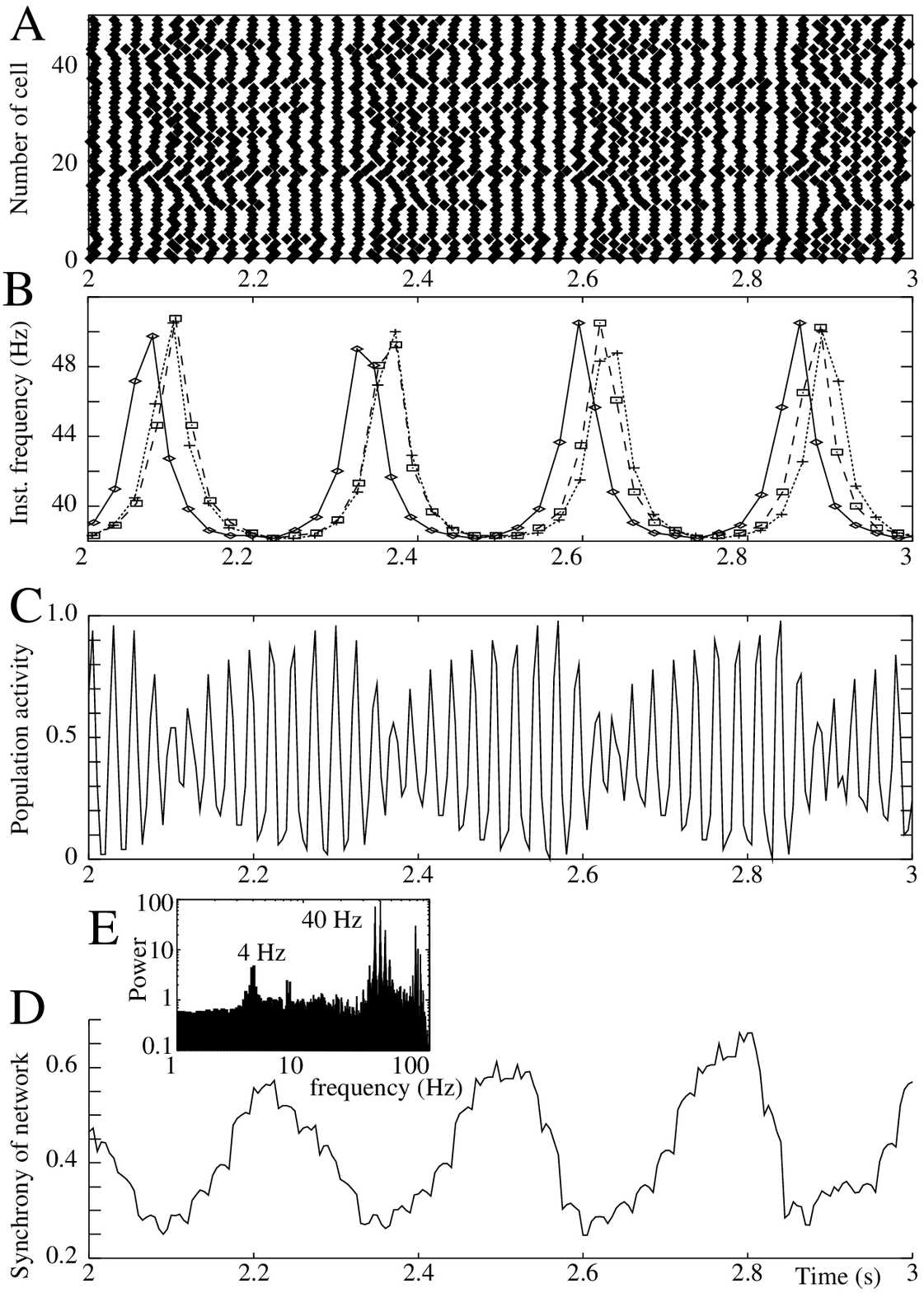}
  \caption[Control activity in the model. (\textit{A}), The raster
  plot shows the firing of all 50 neurons in the network during 1~s.
  When the neurons are synchronized the dots are arranged into
  vertical lines (e.g.\ at 2.8~s). The network looses its synchrony
  when the neurons increase their firing rate (e.g.\ at 2.6~s the
  instantaneous frequency on \textit{B} is high and the raster plot is
  disorganized).  (\textit{B}), Instantaneous frequency of the neurons
  are in the gamma band (40-48~Hz) modulated with a slower oscillation
  (4~Hz, three example neurons are shown). (\textit{C}), The
  population activity shows the proportion of neurons that fired in a
  5~ms long time bin. The amplitude of the oscillation is modulated by
  theta. (\textit{D}), The synchrony of the network is computed from
  the correlation of the activity of the cell-pairs in a 50~ms long
  time bin. (\textit{E}), The Fourier spectrum of the population
  activity shows distinct peaks at 40~Hz (gamma frequency) and at 4~Hz
  (theta frequency). Note that the first second, the transient
  behavior of the model was omitted. Parameters used (see Appendix):
  number of cells: 50, $\bar{g}_\text{syn}=$ 0.2~mS/cm$^2$, connection
  probability: 0.8, $\omega=$ 37~Hz.]{}
  \label{fig:control}
\end{figure}

As argued in \cite{kiss01} this synchronization\,--\,desynchronization
of the population activity of interneurons is transposed to principal
cells of the hippocampal CA3 modulating their firing probability in
the theta band. This modulation might be reflected in the EEG or CA3
local field potential. Thus, we propose that there is a mapping
between the FFT of the CA3 local field potential or EEG and the FFT of
the simulated population activity (Fig.~\ref{fig:control}\textit{E}),
which shows a peak at the theta frequency. Other peaks at higher
frequencies are related to the fast firing properties and the
resulting synchronization of the interneurons as well as the frequency
of the driving current.

\textbf{Drug administration:} Given the detailed kinetic description
of GABA$_\text{A}$ receptors the modulation of the synaptic current
due to its interaction with the drug can be simulated.  Moreover, in
the integrated model framework an evaluation of the drug effect on the
network level can be performed.

The identification of the effect of a given drug on the kinetic rate
parameters of a certain receptor type or subtype is a complicated and
tedious undertaking. To present the soundness of our model two known
drugs were used in the simulations: propofol and midazolam (for
parameter values see the Appendix). However, a considerable effort was
made to estimate the change of the rate constants due to the
application of the \mbox{L838,417} compound using the chemical method
of parameter estimation. Unfortunately, sufficient kinetic or even
pharmacodynamic data could not be found in the public domain to obtain
realistic estimation for the rate constants.

\begin{figure}
  \includegraphics{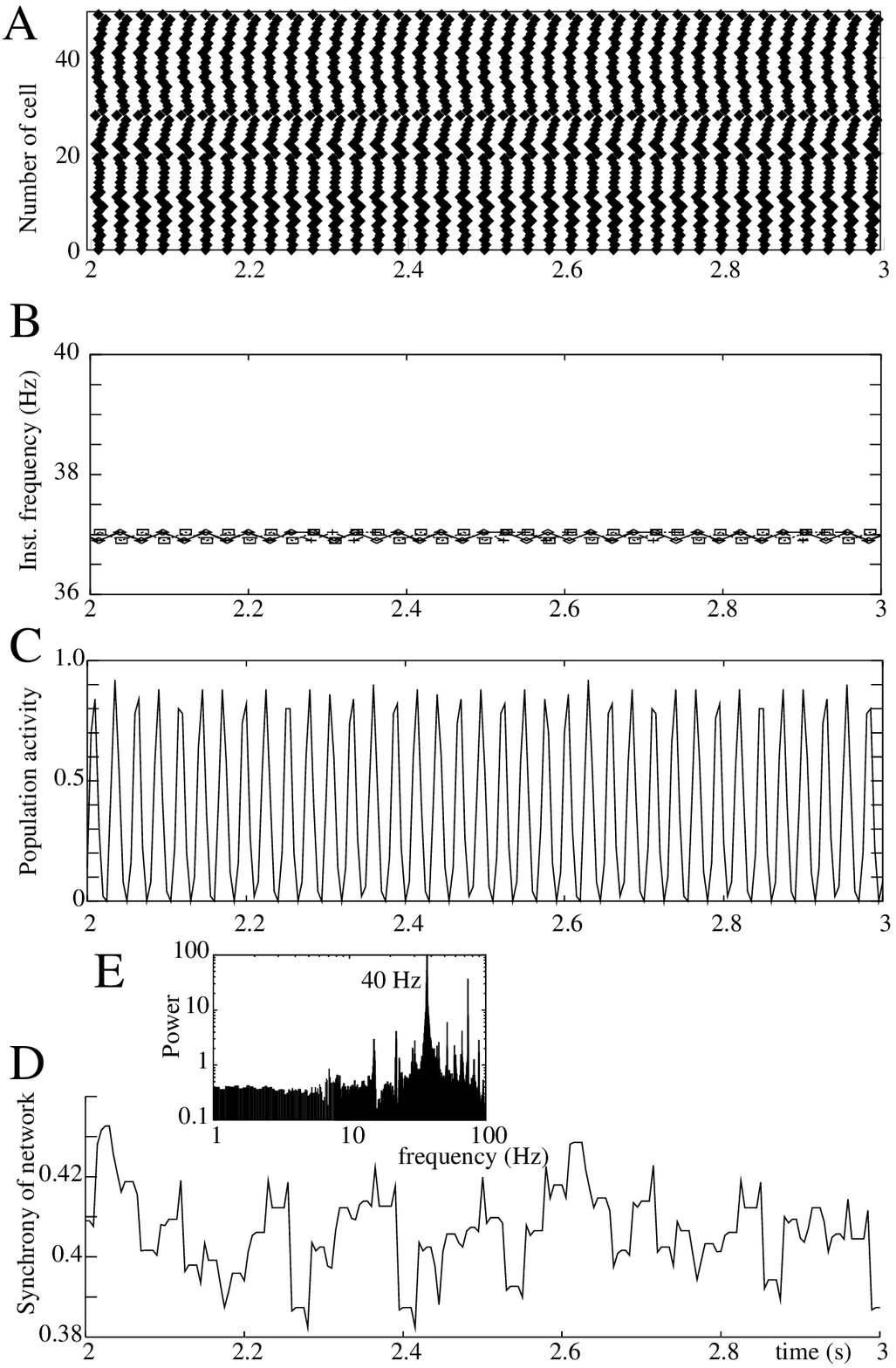}
  \caption[The effect of propofol on the activity of the modeled
  network. The raster plot (\textit{A}) and the population activity of
  the network (\textit{C}) show clear gamma synchrony without any
  slower modulation. The instantaneous frequency of the neurons is
  nearly constant (\textit{B}). Changes in the synchrony of the
  network (\textit{D}) are one order smaller than in control
  condition.  The only peak in the power spectrum of the population
  activity (\textit{E}) is at the gamma frequency. Parameters are
  the same as for Fig.~\ref{fig:control}.]{}
  \label{fig:propofol}
\end{figure}

\begin{figure}
  \includegraphics{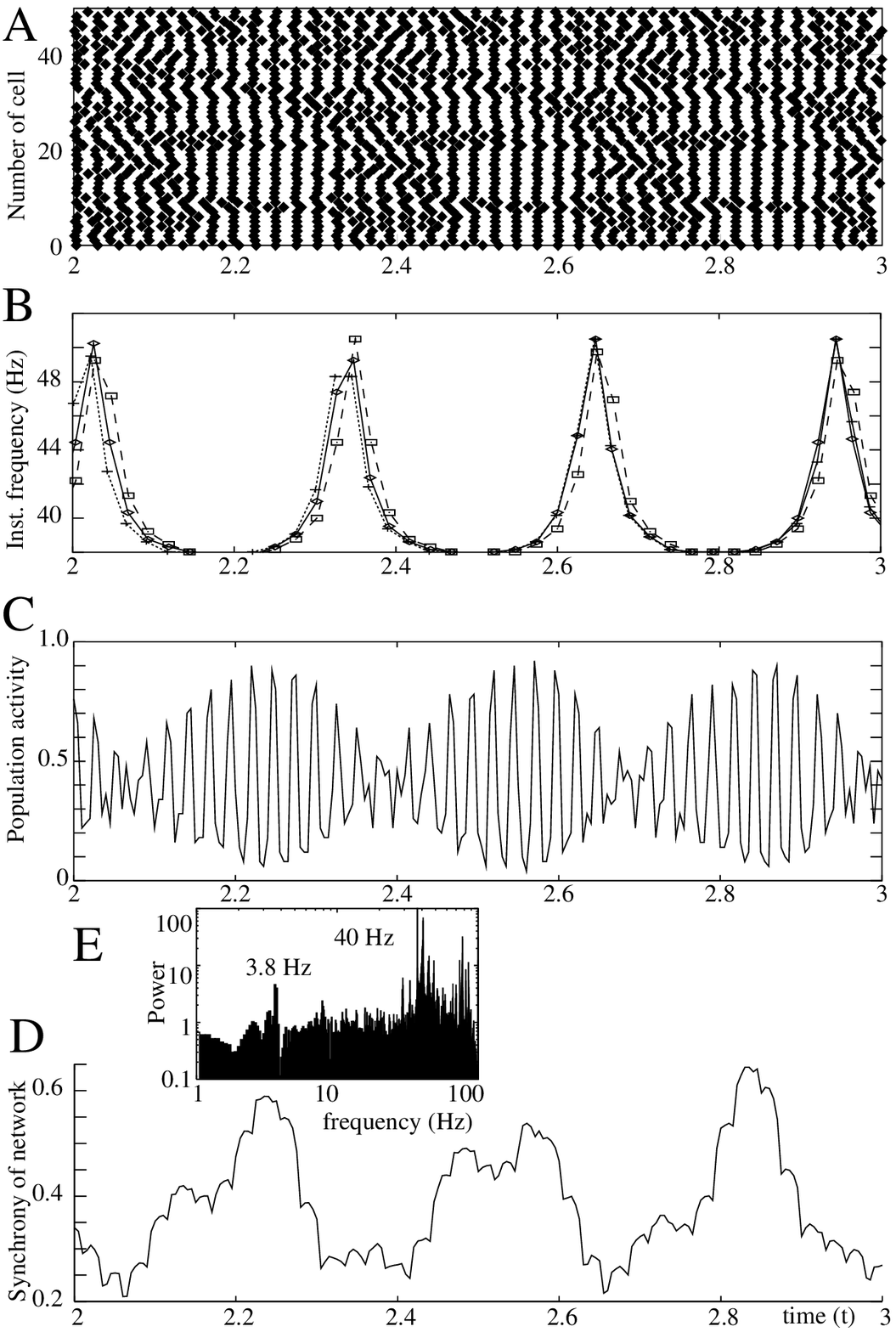}
  \caption[The effect of midazolam on the network activity. Similarly
  to the control condition activity of the neurons are modulated with
  a slower, theta frequency oscillation under the effect of midazolam
  as shown by the the raster plot (\textit{A}); the instantaneous
  frequency of the neurons (\textit{B}); the population activity
  (\textit{C}); the synchrony of the network (\textit{D}) and the
  Fourier spectrum (\textit{E}). Parameters are the same as for
  Fig.~\ref{fig:control}.]{}
  \label{fig:midazolam}
\end{figure}

Figure \ref{fig:propofol} and \ref{fig:midazolam} show effects of
propofol and midazolam, respectively, on the network behavior
(sub-figures are the same as in Fig.\ \ref{fig:control}: \textit{A}:
raster plot; \textit{B}: the instantaneous frequency versus time;
\textit{C}: the population activity versus time; \textit{D}: network
synchrony; \textit{E}: FFT of the population activity function). The
purpose of the present paper is not to give a detailed description and
analysis of the mode of action of these drugs, thus we restrict
ourselves to mention that while propofol prolongs deactivation of the
receptor and reduces the development of both fast and slow
desensitization, midazolam facilitates the accumulation of the
receptors into the slow desensitized state, which compensates the
current resulting from slower deactivation \cite{bai}.  As a result
propofol enhances the tonic synaptic current to a greater extent than
midazolam, whereas propofol and midazolam produces similar changes to
the time course of single inhibitory postsynaptic currents
(IPSCs) \cite{bai98}.

On the network level the difference that propofol and midazolam exert
on the tonic component of the synaptic current has a grave
consequence. While the peak in the FFT at approximately 4~Hz
representing the theta modulation of the population activity is still
present in the case when synapses were ``treated'' with midazolam
(Fig.\, \ref{fig:midazolam}\textit{E}) it is completely missing for
the propofol ``treated'' case (Fig.\, \ref{fig:propofol}\textit{E}).

These findings are in concert with our earlier results \cite{hajos04}
where we have shown that diazepam a positive allosteric modulator of
the \GA synapses inhibited theta oscillation. There synapses were
modeled using the phenomenological technique and the increase of the
tonic component of the synaptic current was achieved by increasing the
maximal synaptic conductance parameter, while the time constants were
kept constant to account for similar IPSCs. Methodologically, the
approach we followed in the integrated model is more established, as
generally the maximal synaptic conductance parameter is mostly
associated with channel density in the synapse, which is not modified
by the administration of the drug. Moreover, using the detailed
kinetic model of the \GA synapse allows us to \textit{i}, account for
drug induced changes in the receptor in a chemically more sound way;
\textit{ii}, simulate a much larger variety of situations than by
using the phenomenological model; and \textit{iii}, present several
possibilities (several sets of rate constants) resulting in the same
receptor behavior (a given receptor behavior might be brought about by
a non-unique set of parameters) to drug researchers who in turn can
engineer the desired molecule that is easiest to design.

\section{Discussion}
\label{sec:discu}

\subsection{Design (Inverse Problem):  From System Identification to
  Optimal Temporal Patterns}

We have shown that in a moderately complex conductance-based model of
the hippocampal CA3 region theta rhythm generation can be observed and
its general properties can be identified \cite{orban01} These results
qualify the model for consideration as a useful tool in the hands of
pharmacologists, physiologists and computational neuroscientists to
complete their repertoire of available tools in the search for
efficient and specific drugs.

\begin{figure}
  \hspace{-2cm}\includegraphics[width=1.4\textwidth]{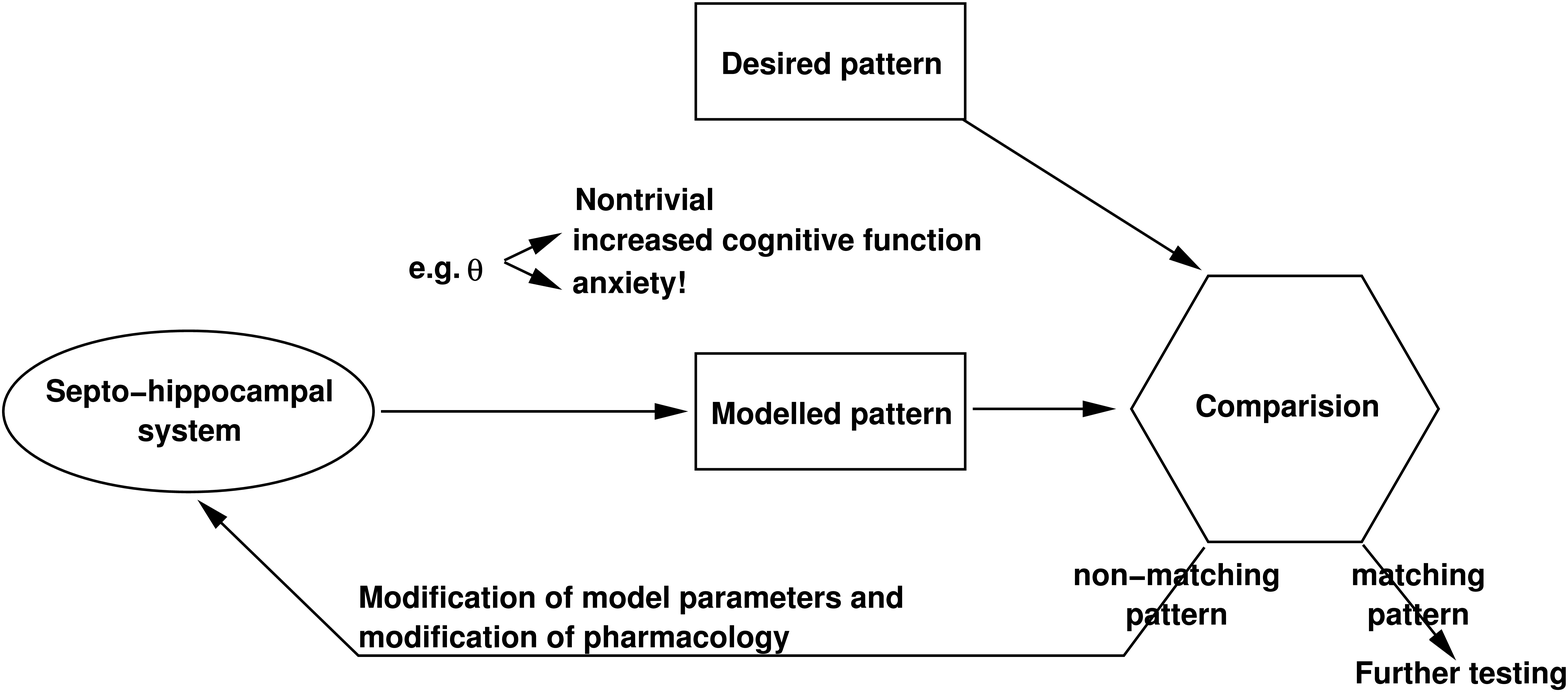}
  \caption[Computational neuropharmacology---an idealized method for
  drug discovery. See text for a description]{}
  \label{fig:compfair}
\end{figure}

Figure \ref{fig:compfair} is an oversimplified scheme offered for
finding a modulator to set optimal septo-hippocampal EEG pattern.

In order to decrease anxiety first a desired EEG pattern should be
defined. Anxyolitics should reduce the theta amplitude (but preserve
the cognitive performance and avoiding sedative and hypnotic side
effects). Computational analysis should offer a kinetic scheme and the
best rate constants to modulate the fixed network to minimize the
deviation from the desired ``optimal pattern''. (Network architecture
is supposed to be fixed. By neglecting this assumption we should turn
from neuropharmacology to neurosurgery...)  Most likely there are more
than one possibilities to reach the goal, and model discrimination and
parameter estimation techniques may help to narrowing the
alternatives.

As it is known from chemical kinetics \cite{et,turtamas90} sensitivity
analysis shows that in a kinetic scheme there are ``more and less
important'' components and reactions. It may help to answer the
question how to modify a given drug to change the reaction rate
constants in the desired direction -- and leaving everything else
intact.

\subsection{Further research}

The aim of the present paper is to offer conceptual and mathematical
frameworks to integrate network and receptor level descriptions for
investigating the effects of potential drugs on the global electrical
patterns of a neural center, and on the behavioral states (mood,
consciousness, etc.). Once we have understood (i) the basic mechanisms
of rhythm generation, (ii) the elementary steps of the modulatory
process, we shall be able to give advice to drug designers pointing
out which subprocess and how to be modulated to reach the given goal.

Specifically, we briefly reviewed some aspects of \GA receptor
kinetics, and the effects of (full and partial) agonists, antagonists
and inverse antagonists to septo-hippocampal theta rhythms. Our
specific goal is to offer a computational platform to design
anxiolytic drugs with as small as possible side effects.  While it is
known that positive allosteric modulators acting on \GA $\alpha_1$
subunits are potential candidates for being selective anxyolitics,
integrative computational modeling would help to select potential
drugs with the appropriate kinetic properties.

\section{Acknowledgments}
Thanks for the motivation and experimental data to Mih\'aly Haj\'os.
We benefited from discussions with Ildik\'o Aradi (PE), Jean-Pierre
Rospars (JT) Gerg\H{o} Orb\'an (TK) and Shanna G.\ Barkume (PE, LZ,
TK).  This work was supported by the Hungarian National Science
Foundation (OTKA) Grant number T038140.  PE thanks the Henry R. Luce
Foundation the general support.

\newpage
\renewcommand{\theequation}{A-\arabic{equation}}
\setcounter{equation}{0}
\section*{APPENDIX}

\subsection*{Interneuron model}

The interneuron model was taken from \cite{wang96} and obeys the
following current balance equation:

\begin{subequations}
\begin{align}
  C_{\text{\tiny{m}}}\frac{\text{d}V}{\text{d}t}&=
  -I_\text{Na}-I_\text{K}-I_\text{L}-I_\text{syn}+I_i\label{eq:dv}\\
I_\text{L}&=g_\text{L}\left(V-E_\text{L}\right)\label{eq:leak}\\
I_\text{Na}&=g_\text{Na}m_{\infty}h\left(V-E_\text{Na}\right)\label{eq:Ina}\\
m_{\infty}&=\frac{{\alpha}_\text{m}}{{\alpha}_\text{m}-{\beta}_\text{m}}\\
{\alpha}_\text{m}&=\frac{-0.1\left(V+35\right)}{{\exp}\left(-0.1
\left(V+35\right)\right)-1}\\
{\beta}_\text{m}&=\frac{1}{{\exp}\left(-0.1\left(V+28\right)\right)}\\
\frac{\mathrm{d}h}{\mathrm{d}t}&={\phi}\left({\alpha}_\text{h}\left(1-h\right)-
{\beta}_\text{h}h\right)\\
{\alpha}_\text{h}&=0.07\:{\exp}\left(\frac{-\left(V+58\right)}{20}\right)\\
{\beta}_\text{h}&=\frac{1}{{\exp}\left(-0.1\left(V+28\right)\right)}\\
I_\text{K}&=g_\text{K}n^4\left(V-E_\text{K}\right)\\
\frac{\mathrm{d}n}{\mathrm{d}t}&={\phi}\left({\alpha}_\text{n}
\left(1-n\right)-{\beta}_\text{n}n\right)\\
{\alpha}_\text{n}&=\frac{-0.1\left(V+34\right)}{{\exp}
\left(-0.1\left(V+34\right)\right)-1}\\
{\beta}_\text{n}&=0.125\:{\exp}\left(\frac{-\left(V+44\right)}
{80}\right)\label{eq:betan}
\end{align}
\end{subequations}

\noindent with parameters: $g_\text{L}=1$\,S/m$^2$,
$g_\text{Na}=350$\,S/m$^2$, $g_\text{K}=90$\,S/m$^2$,
$E_\text{L}=-65.3$\,mV, $E_\text{Na}=55$\,mV, $E_\text{K}=-90$\,mV
$\phi=5$,

\subsection*{Synapse model}

The presynaptic model describes transmitter release due to presynaptic
action potentials. The following sigmoid function was used:

\begin{equation}
  [L](t) = \frac{0.003}{1+\exp\left(-\frac{V_\text{pre}(t)}
      {2}\right)}\label{eq:presyn}
\end{equation}

\noindent where $[L](t)$ is the released GABA, $V_\text{pre}(t)$ is the
presynaptic membrane potential.

Our starting point for the postsynaptic model is the work of Bai et
al. \cite{bai} originally developed by Celentano and Wong \cite{cw} to
describe the emergence of open channels as the effect of GABA binding
to receptors.  However, we explicitly show the presence of an
essential participant of the model, the ligand $L$ differently from
in Scheme 1 of the cited paper and our previous work \cite{et}.
Obviously, in the model the ligand $L$ denotes GABA, whereas
$L_2O$ denotes the open channels, the notation expressing the fact
that two ligand molecules are needed to change the receptor into an
open channel:

\singlespacing
\begin{eqnarray}
\begin{array}{cccccccc}
&&&&&{\ }{\ }{\ }{\ }{\ }L_2D_{\textrm{fast}}&&\\
\\
&&&&&d_f \upharpoonleft\negthickspace\downharpoonright r_f&&\\
&2k_{\textrm{on}}&&&k_{\textrm{on}}&&d_s&\\
L+C&\rightleftharpoons&L_1C{\ }&{\ }L+L_1C&\rightleftharpoons&L_2C&\rightleftharpoons&L_2D_{\textrm{slow}}\\
&k_{\textrm{off}}&&&2k_{\textrm{off}}&&r_s&\\
&&&&&\alpha\upharpoonleft\negthickspace\downharpoonright \beta&&\\
\\
&&&&&L_2O&&\\
\end{array}\label{basicmodel}
\end{eqnarray}
\doublespacing

The induced differential equations are the followings:

\begin{subequations}
\begin{align}
{[C]'}&=-2k_{\textrm{on}}[L][C]+k_{\textrm{off}}[L_1C]\label{dec}\\
{[L_1C]'}&=2k_{\textrm{on}}[L][C]+2k_{\textrm{off}}[L_2C]-(k_{\textrm{off}}+k_{\textrm{on}}[L])[L_1C]\label{del1c}\\
{[L_2C]'}&=k_{\textrm{on}}[L][L_1C]+r_f[L_2D_{\textrm{fast}}]+r_s[L_2D_{\textrm{slow}}]+\alpha[L_2O]\\
&-(2k_{\textrm{off}}+d_f+d_s+\beta)[L_2C]\label{del2c}\\
{[L_2D_{\textrm{fast}}]'}&=d_f[L_2C]-r_f[L_2D_{\textrm{fast}}]\label{del2df}\\
{[L_2D_{\textrm{slow}}]'}&=d_s[L_2C]-r_s[L_2D_{\textrm{slow}}]\label{del2ds}\\
{[L_2O]'}&=\beta[L_2C]-\alpha[L_2O].\label{del2o}
\end{align}
\end{subequations}

Furthermore, we have the following conservation equation for the
total quantity of channels in any form:
\begin{equation}
[C]+[L_1C]+[L_2C]+[L_2D_{\textrm{fast}}]+[L_2D_{\textrm{slow}}]+[L_2O]=[C](0)=1.
\label{conserv}
\end{equation}

Rate constants in the model were set up according to
\cite{bakeretal02} to account for the control situation, the propofol
treatment and the midazolam treatment:

\singlespace
\begin{tabular}{llll}
& \multicolumn{3}{c}{Rate Constants (ms$^{-1}$)}\\
Parameters & Control & Propofol & Midazolam \\
\hline
$k_\text{on}$ & 1000/M & 1000/M & 1000/M\\
$k_\text{off}$ & 0.103 & 0.056 & 0.056\\
$d_\text{f}$ & 3.0 & 1.62 & 3.0\\
$r_\text{f}$ & 0.2 & 0.12 & 0.2\\
$d_\text{s}$ & 0.026 & 0.014 & 0.026\\
$r_\text{s}$ & 0.0001 & 0.0001 & 0.0001\\
$\alpha$ & 0.4 & 0.4 & 0.4\\
$\beta$ & 6.0 & 6.0 & 6.0
\end{tabular}

\doublespace

\subsection*{Data analysis}

\subsubsection*{The raster plot}

A raster plot is used to visualize spike-trains of neurons. To draw a
raster plot, first spiking times of neurons were identified, then time
was discretized assigning a 0 or 1 value to each time bin depending on
weather the cell emitted a spike in a given bin or not:

\begin{subequations}
\begin{align}
  t_i^\text{spiking} &= \{t \,\,|\,\, V_i(t) > 0 \wedge
  \dot{V}_i(t) \ge 0\}\\
  F_i^\tau(l) &= \left|t_i^\text{spiking} \cap \{t \,\,|\,\, l\tau \le
    t < (1+l)\tau\}\right|
\end{align}
\end{subequations}

\noindent where $t_i^\text{spiking}$ is the set of time points when
cell $i$ emits an action potential, $V_i(t)$ is the membrane potential
of cell $i$, $F_i^\tau(l)$ is the spike train and $\tau=5$~ms is the
binning width.

\subsubsection*{The instantaneous frequency}

To characterize the behavior of individual cells a measure of their
instantaneous frequency or firing rate was calculated the difference as in
\cite{orban01}:

\begin{subequations}
\begin{align}
  \mathrm{ISI}_i(t) &=\min\left(t_i^\text{spiking}\cap{t_2|t_2>t}\right)
  -\max\left(t_i^\text{spiking}\cap{t_1|t_1\le t}\right)\\
  f_i(t) &=\frac{1}{\mathrm{ISI}_i(t)}
\end{align}
\end{subequations}

\subsubsection*{The population activity measure}

A ``projection'' of the raster plot to the time axis gives the
population activity:

\begin{equation}
  a^{\tau,T'}(t) =
  \frac{1}{N}\sum_{i=1}^N\sum_{l=\left[\frac{t}{\tau}\right]}^
    {\left[\frac{t+T'}{\tau}\right]} F_i^\tau(l)
\end{equation}

\noindent where $N$ is the number of neurons, $T'=10$~ms is a time
window in which action potentials are counted and $[m]$ denotes the
integer part of $m$.

\subsubsection*{The network synchrony}

The network synchrony measure was defined as follows:

\begin{equation}
  \kappa^{\tau,T'}(t)=\frac{2}{N(N-1)}\sum_{i=2}^{N}\sum_{j=1}^{i-1}\frac{\displaystyle\sum_{l=\left[\frac{t}{\tau}\right]}^{\left[\frac{t+T'}{\tau}\right]}F^\tau_i(l)F^\tau_j(l)}{\sqrt{\displaystyle\sum_{l=\left[\frac{t}{\tau}\right]}^{\left[\frac{t+T'}{\tau}\right]}(F_i^\tau)^2(l)\displaystyle\sum_{l=\left[\frac{t}{\tau}\right]}^{\left[\frac{t+T'}{\tau}\right]}(F_j^\tau)^2(l)}}
\end{equation}

where the window width was set to $T'=50$~ms.

\end{document}